# Spontaneous Emission of Vector Vortex Beams

Domitille Schanne[1], Stéphan Suffit[1], Pascal Filloux[1], Emmanuel Lhuillier[2], Aloyse Degiron[1,*]

[1]Université de Paris, CNRS, Laboratoire Matériaux et Phénomènes Quantiques, F-75205 Paris, France

[2]Sorbonne Université, CNRS, Institut des NanoSciences de Paris, INSP, F-75005 Paris, France

ABSTRACT: Harnessing the spontaneous emission of incoherent quantum emitters is one of the hallmarks of nano-optics. Yet, an enduring challenge remains—making them emit vector beams, which are complex forms of light associated with fruitful developments in fluorescence imaging, optical trapping and high-speed telecommunications. Vector beams are characterized by spatially varying polarization states whose construction requires coherence properties that are typically possessed by lasers—but not by photons produced by spontaneous emission. Here, we show a route to weave the spontaneous emission of an ensemble of colloidal quantum dots into vector beams. To this end, we use holographic nanostructures that impart the necessary spatial coherence, polarization and topological properties to the light originating from the emitters. We focus our demonstration on vector vortex beams, which are chiral vector beams carrying non-zero orbital angular momentum, and argue that our approach can be extended to other forms of vectorial light.

I. INTRODUCTION

Vector beams are optical modes with a spatially inhomogeneous field structure. Cylindrical vector beams, for example, are characterized by polarization singularities that can be produced with birefringent crystals, polarizing prisms, or nanostructured media [1,2]. These elements are either directly integrated within a laser cavity or used as external passive elements in free space. Vector vortex beams [3–6], for their part, combine the polarization singularities of cylindrical vector beams and the properties of scalar vortices [7,8], namely, a phase singularity caused by chiral helical wavefronts and a non-zero orbital angular momentum. Vector vortex beams are often generated by making a laser interact with cascading and/or interfering elements so as to not only generate a non-trivial polarization state, but also impart a non-zero orbital angular momentum to the mode. As for scalar vortices, this latter condition is usually fulfilled with phase plates [8], q plates [3,9], phase holograms [10], or optical metasurfaces [4,11].

The many choices at hand to produce such forms of complex light are a consequence of two basic properties of laser radiation. First, the polarization state of the laser provides a valuable degree of liberty, as many of the aforementioned approaches only work for specific polarizations (e.g. circularly polarized modes). Second, the high spatial coherence of lasers allow them to interact with a variety of structures regardless of the length scales involved in the formation of the beam.

When placed in a homogeneous environment, the emission of light by an excited atom, molecule or quantum dot is random, unpolarized and incoherent. However, it was recognized decades ago that the fluorescence of organic emitters could be rendered linearly polarized and directional with surface plasmons [12–14], which are electromagnetic modes guided at the surface of metals. Although not explained in such terms at the time, these results are related to the coherence lengths of surface plasmons. Rather than emitting light in free space, the fluorophores of these experiments excite surface plasmon modes with spatial correlations that exceed several wavelengths, allowing them to leak light into polarized narrow lobes if the metallic structure is properly designed (e.g. using a Kretschmann-Reather configuration [12] or by carving a diffraction grating [13,14] at the metal surface).

An impressive number of studies has since built on these pioneering results [15–23]. Problems that were long deemed extremely challenging have been addressed one by one, including shaping Planck's blackbody radiation [15,21], beaming light from individual quantum emitters [17,18], and, very recently, generating circularly polarized single photons [23]. These developments have remained focused on the spontaneous emission of conventional scalar modes into controlled directions of space, even if it is worth noting that the light thus produced can be subsequently converted into more complex beams using far-field optical elements [24]. In this article, we go a step further and show how to make an assembly of colloidal quantum dots spontaneously emit optical vector vortex beams.

II. RESULTS

We consider the situation depicted in the inset of Fig. 1(a): a compact layer of PbS colloidal quantum dots (CQDs) coated on top of a structured Au plane. The CQDs have an average diameter of 5-6 nm, they are capped by oleic acid and emit in the near-infrared range [Fig. 1(b)]. At the basis of our design strategy is the hypothesis that the CQDs are not quenched by the presence of the metal. Furthermore, we posit that their photoluminescence predominantly excites the plasmon modes of the system (these assumptions will be justified later). When the sample is pumped by a laser focused on the surface, the area of excited dots can be considered in a first approximation as a point source of plasmons that propagate radially in all directions along the plane of the sample [Eq. (1) of the Appendix].

To convert these plasmons into light, we decorate the Au surface with a grating to make them diffract into free space. Different from the periodic patterns usually chosen to accomplish this function [13,14], we imprint a holographic topological defect into the grating to generate a vector vortex beam with a $m^{th}$ order phase singularity at its center, m being the topological charge of the mode [Eq. (2) of the Appendix]. The resulting pattern is an m-arm Archimedean spiral, which is a variant of the celebrated fork hologram used to generate optical vortices with free-space Gaussian laser beams [10]. To predict the intensity and phase distribution of the beam resulting from the radiative decoupling of the surface plasmons, we calculate the far-field diffraction pattern of the leaked radiation in the scalar Fraunhofer approximation [Eq. (3) of the Appendix].

Figure 1(a) shows a scanning electron micrograph of a structure designed to launch a vector vortex beam with topological charge m=3. The total radius of the structure is 100 µm, the spiraling grooves are 25 nm thick and the distance between two adjacent grooves (radial pitch) is 1.095 µm. Although not visible at this magnification, the spiral is coated by a continuous carpet of two to three monolayers of PbS CQDs. The sample is characterized with a custom upright microscope. To excite the CQDs, a 633 nm HeNe laser is focused with a 50X objective on the surface of the sample. The photoluminescence is collected through the same objective, separated from the pump using a dichroic mirror, and sent on an imaging spectrograph coupled to an InGaAs Camera. An additional Bertrand lens is inserted into the optical path to image the back focal plane of the objective, thus revealing the transverse angular distribution (or, equivalently, the in-plane wavevectors $k_x$ and $k_y$) of the photons emitted by the sample [25].

We first perform a series of experiments in which the focused laser spot excites the CQDs at the center of the m=3 spiral. Figure 1(c) shows the angular distribution of the photoluminescence at three representative wavelengths λ, obtained by imaging the back focal plane of the objective through different narrow bandpass filters. These images are displayed in normalized units and the relative intensities will be discussed later. The experimental results are in good agreement with the predictions of the model displayed in Fig. 1(d). For all photoluminescence wavelengths, a well-defined beam against a weak isotropic background is observed. Furthermore, the angular distribution of the beam evolves from a large ring at 1050 nm to a small spot at 1100 nm before becoming annular again and increasing in diameter at larger wavelengths. This evolution reflects the dispersive nature of our design approach based on the diffraction of surface plasmons, the in-plane wavevectors being fixed by the conservation of momentum $\mathbf{k_x} + \mathbf{k_y} = \mathbf{k_{SP}} - \mathbf{G}$, where $\mathbf{k_{SP}}$ is the plasmon wavector and $\mathbf{G}$ the reciprocal vector corresponding to the radial periodicity.

At 1100 nm, the calculations predict that the ring becomes very narrow and thus well collimated; however, the latter does not merge into a spot contrarily to what is observed on the experimental image. This small discrepancy arises because the model assumes a point source of radial plasmons while the actual laser pump excites a finite area of CQDs. By refining the model to take this finite extent into account [Eq. (4) of the Appendix], the calculations become fully consistent with the experiments, as shown on Fig. S1 [26].

Figure 1(e) represents the calculated phase for the three wavelengths under investigation: in all cases, the phase is singular at the center of the beam, which is the distinct signature of an optical vortex with non-zero orbital angular momentum. To verify that the experimental phase is consistent with these calculations, we take advantage of the fact that two interfering vortices produce unique signatures that reveal their phase structure and topological charge [8]. Our next experiment has thus been to fabricate and study a sample combining two spiraling gratings with opposite windings, so as to simultaneously emit the same beam as before (m=3) together with another vortex with opposite topological charge n=-3 [Fig. 2(a)]. The resulting emission pattern, displayed in Fig. 2(b), features six star-shaped branches that are well reproduced by our model that treats the photoluminescence of the CQDs as a point-like source of radial waves [Fig. 2(c)]. It is even possible to achieve a near-perfect agreement with the measurements by taking into account the fact that the laser pump excites an area of CQDs that does not reduce to a single point, as shown on Fig. 2(d) where we have evaluated and summed the diffraction patterns at multiple locations in a region around the hologram center using Eq. (4) of the Appendix.

The other panels of Fig. 2 suggest that these results can be generalized to arbitrary combinations of vortices such as (m,n) = (4,-4), (5,-5), and (5,0). The experimental patterns always exhibit |m-n| arms and are in very good agreement with the calculations, indicating that the beams have the anticipated topological charges and phase singularities of genuine vortices.

We now examine the polarization state of the beams. In Fig. 3, we have repeated the experiments shown in Fig. 1 with the same m=3 structure, but with a linear polarizer inserted between the sample and the InGaAs camera. The different images depict the evolution of the emission as the polarizer axis is rotated. In all cases, the recorded images feature two lobes oriented perpendicular to the axis of the polarizer. In other words, the polarization is not homogeneous in the plane transverse to the propagation but exhibits the characteristics of a radial vector beam. We note, however, that the intensity of the beam does not strictly fall to zero in the direction perpendicular to the maxima. Thus, the purity of the vector vortex beam is good, but not perfect, a result consistent with the fact that the structure has been designed assuming a single point-source at its center while we have seen previously that a finite area of CQDs contributes to the emission. This being said, we can say with confidence that the emission is not an equal mixture of vortices with topological charges ±|m| since we would have otherwise observed the distinctive interferometric patterns evidenced in Fig. 2.

The results of Fig. 3, shown here only for a three-arm spiral at 1050 nm, hold true for all wavelengths and all the m-arm spirals that we have tested. This observation can be understood by having in mind that the diffraction of surface plasmons by a grating produces light polarized with the same azimuthal angle than their propagation direction. Since in our case, the plasmons are propagating radially away from the center of the holograms, this radial distribution is also present in the field structure of the emitted light [27,28].

We next study the emission properties of our structures without the bandpass filters that were used so far to examine the beams at selected wavelengths. Figure 4(a) compares the experimental and theoretical dispersion relations of the full polychromatic beam created by the m=3 spiral. The dispersion of the vortex appears as bright branches that fade away as the beam divergence increases. To assess this divergence, we have plotted on Fig. 4(b) a cross-section of the back focal plane for the full polychromatic beam. The full width at half maximum of the divergence is limited to $k_x/k_0 = \pm\, 0.14$, or $\pm\, 8°$, which is quite narrow considering the large emission bandwidth of our CQDs [from less than 950 nm to more than 1250 nm, cf. Fig. 1(b)].

A closer look at the experimental dispersion of Fig. 4(a) reveals a small gap between the upper and lower vortex branches as well as an asymmetric intensity distribution, the maximum of the luminescence intensity being located on top of the low-energy branch. Such features are another evidence of the role played by surface plasmons—they are typically observed in diffractive plasmonic systems [29] and correspond to the formation of two standing waves with different field symmetries at the center of the first Brillouin zone [30]. While otherwise very accurate, our model does not capture this anticrossing because the calculations are based on a scalar description of the surface plasmons. This approximation, which neglects the fact that plasmonic electric fields have a longitudinal component in addition to a transverse one, is valid in most cases—but not when field symmetries play a role, as is the case for the opening of a gap in a dispersion relation [30]. The cost to pay for working with a scalar model is small since most calculated back focal plane images shown in this study are in excellent agreement with the measurements.

## III. DISCUSSION

Combined together, the results of Figs. 1-4(a) demonstrate that holographic structures, which have been long used to convert lasing light into vector vortex beams, can be adapted to perform the same function for the incoherent spontaneous emission of quantum emitters. Despite the similarities with holograms designed to work with lasers, the operating principles of our structures differ in several non-trivial ways. In fact, the metallic patterns play a triple role, explaining the extreme compactness of the structures. First, they ensure that the photoluminescence occurs under the form of radial surface plasmons with long coherence/propagation lengths: from the width of the beams, it is possible to estimate this length as $\lambda/\Delta\theta \approx 30\pm4$ µm. Second, they ensure that the plasmon diffraction to free-space radiation modes is a vector beam with radial polarization. Last, their spiraling geometry imparts the necessary phase singularity and determines the propagation direction and beam divergence.

It is worth noting that our design approach, based on a scalar description of the surface plasmons, solely focus on this third and last aspect (phase singularity, propagation direction and beam divergence). As explained above, the radial polarization can be understood without modelling because it is inherited from the field structure of the radial plasmons involved in the light emission (coupling of surface plasmons is made to TM radiation modes with respect to the gratings). We note in passing that an interesting special case is the holographic pattern obtained with m=0, corresponding to a bull's eye structure rather than a spiral. In this limit, the emitted light has no orbital angular momentum and no phase singularity, but it remains a radial (cylindrical) vector beam with the same polarization behavior as the one evidenced in Fig. 3.

As for the efficient coupling between the emitting layer and the plasmons, we relied on the recent discovery that a carpet of CQDs cannot be modelled as an ensemble of individual emitters [31,32]. Rather, their mutual interactions lead to the formation of electron and holes reservoirs in local thermodynamic equilibrium, as experimentally demonstrated elsewhere for quantum dot LEDs [31] and in the Supplemental Material [26] for the vortex-emitting samples investigated here. Under such conditions, the usual framework to describe the luminescence of individual nanosources, based on the local density of optical states and the Purcell effect, does not apply. In particular, the latter predicts that individual emitters in direct contact with a metallic surface undergoes complete quenching [17], which is not the case here.

To account for the mutual interactions between the CQDs, a statistical treatment of the photoluminescence is relevant. It leads to a local version of Kirchhoff law [32] which states that tailoring the emission of the CQDs is equivalent to tune their absorption cross-section (at the wavelength of emission). With this picture in mind, we can understand why the CQDs are not quenched and why all the features evidenced in our experiments originate from the diffracted plasmons—they are the main electromagnetic modes supported by the system and their high local fields modify the absorption cross-section of the dots—or, equivalently, their spontaneous emission. As a comparison, we show on Fig. 4(c) the angular distribution of the spontaneous emission obtained with a hologram on a glass substrate (i.e., only the corrugations are made of Au). The other parameters are the same as the image at $\lambda=1200$ nm on Fig. 1(c). The differences are spectacular—not only do we observe two annular rings because the system supports plasmons with different indices on the glass side and the CQD/air side, but the contrast between these beams and the background drops at 17%. [see also the cross-sections plotted on Fig. 4(b)]. This observation is consistent with the fact that for a hologram on glass, free-space optical modes other than plasmons are allowed in the immediate vicinity of the CQDs, allowing strong parasitic light to be emitted in addition to the beams of interest.

IV. CONCLUSION AND PERSPECTIVES

In conclusion, we have pushed the control of the spontaneous emission to a complexity usually achieved with lasers. These results point to further developments in the future. For example, the dispersion with the wavelength could be alleviated or constrained by replacing the binary holograms with more elaborated patterns, such as active metasurfaces exploiting the local Kirchhoff law, as the latter makes it possible to tune the color, polarization state and brightness of the luminescence at a subwavelength scale [33,34]. In addition, this study on vector vortex beams shows a path forward to generate other complex beams out of quantum emitters. More generally, these ideas are valid regardless of the way electrons and holes are generated, promising also fruitful possibilities for electroluminescence and thermal emission.

ACKNOWLEDGMENTS

We acknowledge support from the European Research Council grant FORWARD (reference: 771688).

CORRESPONDING AUTHOR

* aloyse.degiron@u-paris.fr

APPENDIX: EXPERIMENTAL DETAILS AND THEORETICAL MODEL

**Chemicals.** Octadecene (ODE, Acros Organics, 90%), Lead oxide (PbO, Strem Chemicals, 99.999 %), Oleic acid (OA, Alfa Aesar 90%), Hexamethyldisilathiane (TMS$_2$S, Sigma Aldrich, synthesis grade), Ethanol (VWR, >99.9%), Toluene (Carlo Erba, >99.8%). Lead oleate Pb(OA)$_2$ 0.1 M: 0.9 g of PbO are mixed in a three neck flask with 40 mL of oleic acid. The flask is degassed under vacuum for one hour. The atmosphere is switched to Ar and the temperature raised to 150 °C for two hours. The final solution is typically clear and yellowish.

**CQD synthesis.** The procedure is inspired from Hines et al. [35]. 12 mL of Pb(OA)$_2$ (0.1 M) are introduced in the round-bottomed flask. The mixture is stirred and degassed under vacuum at 100 °C for 30 min. Then the atmosphere is switched to Ar and the temperature is set at 80 °C. Meanwhile, in the glove box a mixture of 400 µL of TMS$_2$S and 20 mL ODE is prepared in a 20 mL vial. 6 mL of this solution is introduced into a 10 mL syringe. At 80°C, the content of the syringe is rapidly injected, the solution turns immediately dark. After 1.5 min, the heating mantle is removed, and the flask is cooled down with fresh air flux. The reaction mixture is split into two tubes and the particles are precipitated by adding 25 mL of EtOH in each tube. The mixture is centrifuged at 5000 rpm for 5 min. All the solids are redispersed in 7.5 mL of toluene and 1 drop of OA is added. The particles are precipitated by adding 10 mL of EtOH and centrifuged at 5000 rpm. The pellet is dissolved in 10 mL of toluene. The solution is centrifuged at 5000 rpm, and the formed pellet, containing the non-colloidally stable part is discarded. We reprecipitate the nanoparticles by adding a tiny amount of ethanol in a weighted plastic tube. This pellet is then dried in the antichamber of a glovebox, before getting redispersed in toluene and filtered through a PTFE filter (0.22 µm).

**Model.** The CQDs excited by the focused laser pump can be considered as a localized source of radial plasmons. A strong (but experimentally justified) approximation that we make throughout the study is to neglect the vectorial field structure of the plasmons and model their electric field as a scalar cylindrical wave $A$:

$$A(x - x_i, y - y_j) \approx \exp\left[in_{eff}k_0\sqrt{(x-x_i)^2 + (y-y_j)^2}\right]/\left[(x-x_i)^2 + (y-y_j)^2\right]^{1/4}, \quad (1)$$

where $(x_i, y_j)$ are the coordinates of the origin of the radial wave in the $(x,y)$ plane of the sample, $k_0$ is the free space wavevector and $n_{eff} = 1.012+0.003i$ is the measured plasmon effective index (see below). To design the structures, we calculate the hologram resulting from the interference between a reference radial distribution centered at the origin and an exit wave carrying a $m^{th}$ order phase singularity $\exp(im\theta)$:

$$H(x, y, \theta) = \left|\exp\left[iRe\left(n_{eff}\frac{2\pi}{\lambda_c}\sqrt{x^2+y^2}\right)\right] + \exp[im\theta]\right|^2, \quad (2)$$

and convert this smoothly oscillating function of $x$ and $y$ and the polar angle $\theta$ into a binary structure $H_{bin}$ of finite size compatible with e-beam lithography. In this equation, $\lambda_c$ is the wavelength of the collimated vortex (i.e. the one whose angular distribution reduces to a point in the back focal plane images). Finally, we predict the structure of the beam by calculating the far-field diffraction pattern of the leaked radiation in the Fraunhofer approximation:

$$I_{PL}(k_x, k_y) = \left|\mathcal{F}\{H_{bin}(x,y) A(x - x_i, y - y_j)\}\right|^2. \quad (3)$$

In this expression, $\mathcal{F}$ denotes a Fourier transform, while $k_x$ and $k_y$ are the projections of the emitted wavevector along the plane of the sample. The total footprint of $H_{bin}$ is a circular area with a diameter of 100 µm.

For all figures except Figs. 2(d), 2(h), 2(l), 2(p) and S1, we assumed that the surface plasmons originated from a single point source placed in the center of the hologram. In other words, the calculations were performed by taking $x_i = y_j = 0$ in Eq. (3). For Figs. 2(d), 2(h), 2(l), 2(p) and S1, we took into account the fact that the laser pump excites a finite area of the CQD layer around the center of the hologram. To this end, we have evaluated Eq. (3) at multiple locations over a 16 µm² area at the center of the structure. Then, we have weighted each result with a Gaussian approximation of the Airy pattern $G(x_i,y_j) = \exp[-2(x_i^2+y_i^2)/2\sigma^2]$ formed by the laser spot, so as to capture the variations of the pumping power across the section of the laser beam. We chose σ = 3 µm, the red HeNe laser being slightly defocused due to the chromatic aberrations of the microscope objective used to collect the infrared photoluminescence. Finally, the images of Figs. 2(d), 2(h), 2(l), 2(p) and S1 are obtained by calculating the sum of all patterns:

$$I_{PL}(k_x, k_y) = \sum_{x_i,y_j} G(x_i, y_j) \left|\mathcal{F}\{H_{bin}(x,y) A(x - x_i, y - y_j)\}\right|^2. \qquad (4)$$

We have summed the intensities and not the fields because the contribution of each CQD is independent from the other.

**Determination of the plasmon effective index.** To determine the real part of $n_{eff}$, we have compared the experimental data with the results of our model and adjusted Re($n_{eff}$) until the calculations match the measurements. The most effective way to proceed is to fit the dispersion relations [such as the one displayed on Fig. 4(a)] and then verify that the value thus obtained can be used to model all the other data. We found that Re($n_{eff}$) = 1.012 is a good fit for all wavelengths even if real plasmons are in fact slightly dispersive over the spectral range of interest. This value is close but slightly above the analytical index of surface plasmons at an ideal air/Au interface because of the presence of the thin layer of CQDs. To determine Im($n_{eff}$), we have used the experimental linewidths Δθ of the optical vortices and written that the plasmon propagation length is $L_{SP} = 1/[2\text{Im}(n_{eff})k_0] = \lambda/\Delta\theta$. This calculation yields Im($n_{eff}$) = 3×10⁻³, which is slightly larger than for ideal surface plasmons propagating at an ideally flat Air/Au interface at the same wavelengths.

**Mask generation.** The electronic beam lithography (EBL) masks are generated using the Octave-compatible Raith_GDSII Toolbox. First, the analytical expressions of the spiral arms are deduced from Eq. (2) by solving for the maxima of the interference pattern, yielding $r = p\lambda/n_{eff} + m\theta/k_0 n_{eff}$, with $0 \leq p \leq m-1$. Then an Octave script uses this equation to create a binary mask of the spiral. In Figs. 1, 3 and 4, the duty cycle is equal to 50% except at the center of spirals where the different arms are thinned down to avoid overexposure. The interferometric structures of Fig. 2 have been obtained by merging two spiral masks with different topological charges directly on the Raith Pioneer software that pilots our EBL system. The duty cycle of these interferometric samples has been reduced to 30%.

**Sample fabrication.** First, a 200 nm thick Au layer is coated onto a Si substrate using a Plassys MEB 550S electron beam evaporator. The binary holograms are then defined with a sequence of EBL (with a layer of CSAR 62 positive resist from Allresist spin-coated at 3300 rpm onto the sample and the 20kV beam of a PIONEER Two Raith system), Au deposition (25 nm) and overnight lift-off in 2-butanone. Finally, the solution of CQDs is spin-coated onto the sample so as to form a continuous conformal layer above the different holograms. The CQD thickness is 2-3 monolayers according to scanning electron micrographs of cleaved samples. For the sample on glass presented in Figs 4(b)-(c), the fabrication workflow is the same except that the lithography step is directly performed on a borosilicate glass wafer (Plan Optik AG). In this case, a conductive layer (Electra 92, from Allresist) is spun on top of the CSAR 62 resist to evacuate the charges during electron-beam writing.

**Measurements.** Most of the experimental details appear in the text. The setup, shown in Fig. S3 [26], is built around a customized Olympus BX51WI upright microscope equipped with various ports to pump the CQDs and to analyze the resulting photoluminescence. The infrared photoluminescence is separated from the visible pump with a Thorlabs DMLP950R dichroic mirror. The InGaAs camera used throughout the study is a NIRvana 640ST from Princeton

Instruments mounted on an Acton SP-2356 imaging spectrograph from the same manufacturer. For most experiments, the spectrograph is used in imaging mode, with a planar mirror mounted in its turret and an RG780 longpass filter from Melles Griot placed at its entrance to filter the remaining light from the HeNe pump. The back focal plane experiments were realized with a 30 mm Bertrand lens and a 20 mm Telan lens to create an intermediate image in front of the spectrograph. The bandpass filters used to investigate the beams at selected wavelengths are from the Thorlabs "FB" line and have a full width at half maximum of 10 nm. For the dispersion relation of Fig. 4(a), the mirror is replaced by a 85 groove/mm grating to disperse the image of the back focal plane along one direction and a 100 µm slit was inserted at the entrance of the spectrograph to improve the resolution. To obtain the images of Fig. 3, a linear polarizer has been placed in front of the InGaAs camera and the different panels correspond to different orientations of the polarization axis with respect to the sample. All the other experiments of the study have been performed without this linear polarizer. Our samples are fabricated and measured under ambient air. During the first few days, the luminescence spectrum of the dots blue-shifts due to photo-oxidation processes. Although this shift has no influence on the structure of the vortices, all the data presented here have been recorded after stabilization of the photoluminescence spectrum of the CQDs.

FIGURES

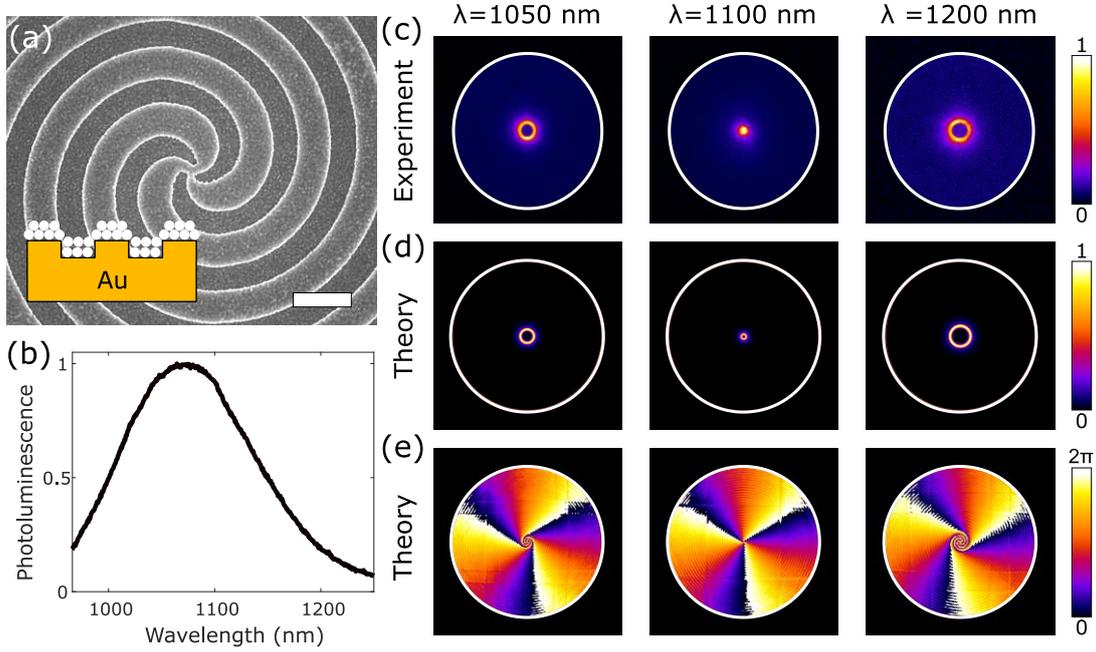

FIG. 1. (a) Scanning electron microscope (SEM) micrograph of an m=3 sample (dimensions given in the text). The light (resp. dark) grey areas are the top (resp. bottom) of the grooves. The vector vortex beam is emitted perpendicularly to this image. Scale bar: 1 μm. Inset: Schematic cross-section of the sample consisting of PbS CQDs (white dots) coated on a Au nanostructured surface. (b) Photoluminescence spectrum of the PbS CQDs. (c) Images in false colors of the back focal plane when the CQDs are pumped with a 633 nm HeNe laser focused at the center of the 3-arm spiral. Each panel corresponds to an observation of the beam profile at a different wavelength, as indicated above the images. The beam propagation is perpendicular to the images. (d) Theoretical predictions of the beam profile at the same wavelengths. (e) Theoretical predictions of the phase. The white circle of all these plots has a radius $k/k_0 = 0.65$, which corresponds to the numerical aperture of our objective.

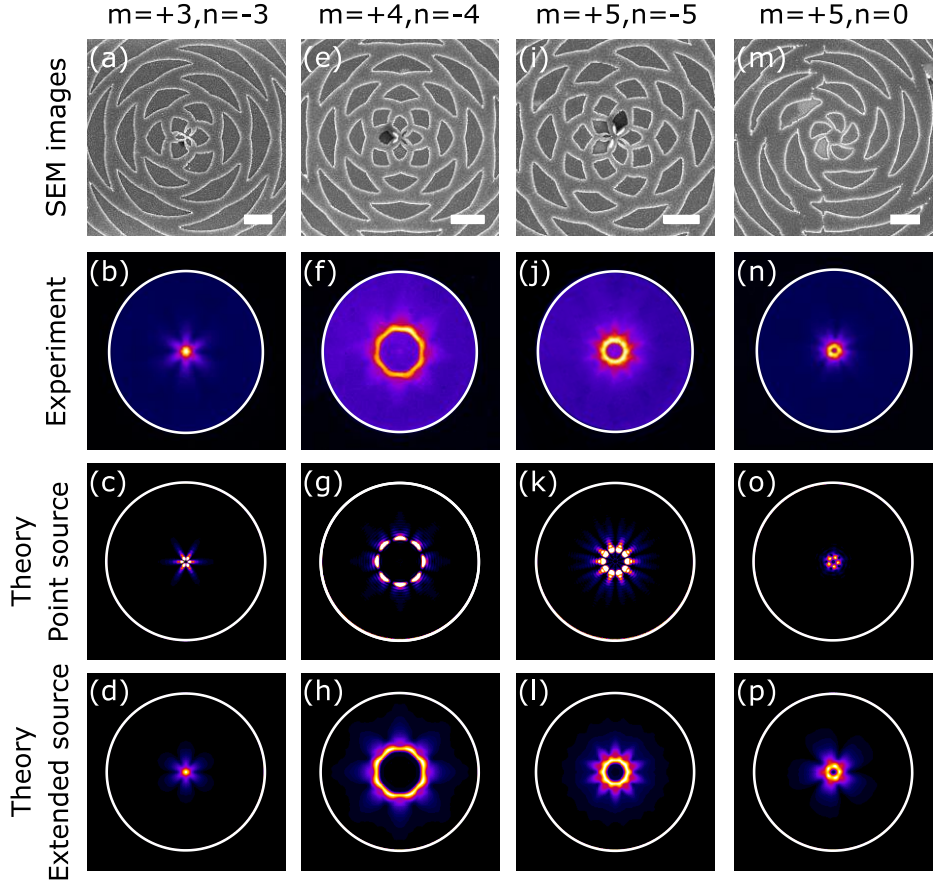

FIG. 2. (a) SEM micrograph of a structure generating two interfering vortices with opposite topological charges m=3 and n=-3 (dimensions given in the text). Scale bar: 1 µm. (b) Experimental image of the back focal plane. (c) Theoretical predictions assuming a point source emitter. (d) Theory when the finite size of the photoluminescence spot is taken into account. (e)-(h) Results for two interfering beams with m=4 and n=-4. The geometry consists of two 4-arm spirals with radial pitch 1.095 µm and opposite windings. (i)-(l) Results for two interfering beams with m=5 and n=-5. The radial pitch of the structure (two 5-arm spirals with opposite windings) is 1 µm. (m)-(p) Results for two interfering beams with m=5 and n=0. Here, the geometry is obtained by superimposing a bull's eye structure with a 5-arm spiral, both structures having a radial pitch of 1.045 µm. The white circle of all these plots has a radius $k/k_0 = 0.65$. The wavelength is 1100 nm for all panels. The color map is the same as in Figs. 1(c)-(d), except that the colors for the point source calculations have been saturated by a few percent for a better comparison with the experiments.

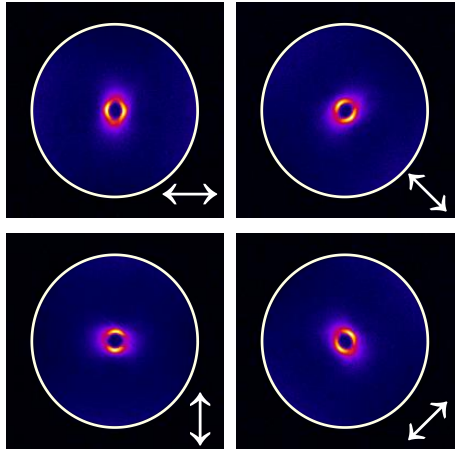

FIG. 3. Images of the back focal plane at 1050 nm for the m=3 structure already studied in Fig. 1 (the same color code has been used). Here, a linear polarizer is inserted in front of the InGaAs camera. The orientation of the polarization axis is indicated with the double arrows on each panel.

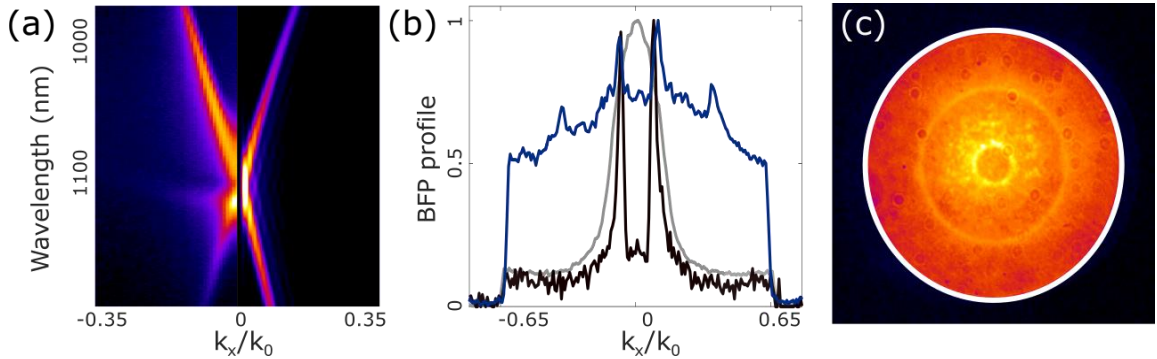

FIG. 4. (a) Experimental (negative $k_x$) and theoretical (positive $k_x$) dispersion relation for the m=3 spiral discussed throughout the text. (b) Normalized back focal plane (BFP) profile of the m=3 beam. The grey curve is the profile of the full polychromatic beam while the black curve is a cross-section of Fig. 1(c) showing the result at the wavelength λ=1200 nm. The blue curve is a cross-section of Fig. 4(c) revealing the profile of the beam at λ=1200 nm produced by a m=3 spiral fabricated on glass (i.e. only the corrugations are made in Au). (c) Same experiment as in the rightmost panel of Fig. 1(c) (λ=1200 nm) for a spiral fabricated on glass. The same color map has been used.